\documentclass{PoS}
\usepackage{floatrow}

\title{Upper limits on gamma-ray emission from Supernovae serendipitously observed with H.E.S.S.}

\ShortTitle{SNe Upper limits with H.E.S.S.}

\author{{\speaker{Rachel Simoni}$^1$, Nigel Maxted$^2$, Mathieu Renaud$^3$, Jacco Vink$^1$, Luigi Tibaldo$^4$ for the H.E.S.S. collaboration}\\
{$^1$GRAPPA/Anton Pannekoek Institute, Universiteit van Amsterdam, Postbus 94249,
1090 GE, Amsterdam, the Netherlands \\ email: { \email{r.c.simoni@uva.nl}} \\
$^2$School of Physics, University of New South Wales, 2052, Sydney, Australia \\email: {\email{n.maxted@unsw.edu.au}}\\
$^3$Laboratoire Univers et Particules, Universit\'e Montpellier, 34095, Montpellier, France }\\
$^4$ Max-Planck-Institut f{\"u}r Kernphysik, P.O. Box 103980, D 69029 Heidelberg, Germany
}

\abstract{Recent theoretical models suggest that young supernovae might be able to accelerate particles, which in turn might generate very high energy gamma-ray emission. We search for gamma-ray emission towards supernovae in nearby galaxies which were serendipitously within the field of view of the High Energy Stereoscopic System (H.E.S.S.) within a year of the supernova event. H.E.S.S. data collected between December 2003 and March 2015 were considered and compared to recent catalogs. Nine candidate supernovae were identified and analysed. No significant emission from these objects has been found. Gamma-ray emission upper limits, which are of the order $\sim$10$^{-13}$\,cm$^{-2}$s$^{-1}$ above 1\,TeV, are reported.}

\FullConference{35th International Cosmic Ray Conference $ - $ICRC2017-\\
		10-20 July, 2017\\
		Bexco, Busan, Korea}

\begin{document}

\section{Introduction}

Although great progress has been made in the search for Galactic cosmic-ray hadron sources up to the knee (1 PeV = ($10^{15}$ eV)) of the cosmic-ray spectrum, we still have not accounted for the bulk of the Galactic cosmic ray flux. Multiple supernova remnants (SNRs) exhibit high-energy and very-high-energy (E>100 GeV) gamma-rays that can be observed by ground-based observatories operating in the TeV energy window. Protons and nuclei accelerated at the SNR shock are responsible for gamma-ray emission in some of these objects (\cite{Ackermann:2013}), while some other objects remain probable particle accelerators (e.g. \cite{Abdalla:2016w49}). These SNRs, which are often $10^{2}-10^{4}$ years old, have never had a presence of significant PeV-energy hadron acceleration demonstrated conclusively, despite this remaining feasible in some sources (e.g. \cite{Fukui:2012}, \cite{Gabici:2014}).

In the search for cosmic-ray sources, young (a few months to a few years) SNe must not be overlooked (e.g. \cite{Cardillo:2015}, \cite{Katz:2011},\cite{Murase:2011},\cite{Marcowith:2014}). Some type IIp, IIb and IIn SNe are embedded in a dense circumstellar medium (CSM) due to the high-mass loss rates and slow wind speeds associated with some progenitor stars (e.g. \cite{Chevalier:review}). In some cases, the resultant SN shock might create a rate of hadron-injection conducive to cosmic-ray shock-acceleration at a time when the SN velocity and density are maximised. Magnetic-field amplification at the shock front and the growth of cosmic-ray streaming-induced plasma instabilities enhance the conditions for shock acceleration.  A population of cosmic ray hadrons from such an environment may leave a signature of gamma-ray emission as p-p interactions occur within the surrounding dense CSM, before heavy CSM-ionisation eases the particle acceleration process. 

The possibility of a brief period of high gamma-ray flux from young SNe offers the H.E.S.S telescope a window to search for signatures of particle acceleration beyond TeV energies. H.E.S.S. data towards 9 serendipitously-observed SNe events were analyzed in a search for gamma-ray emission created by decaying neutral pions in the CSM surrounding young SNe. Upper limits on gamma-ray flux towards these SNe are presented.

\section{Observations and data analysis}
\subsection{H.E.S.S. observations}
H.E.S.S., the High Energy Sterescopic System, is an array of five 
imaging atmospheric Cherenkov telescopes (IACTs) located in the Khomas Highland of Namibia at an altitude of 1800\,m above sea level. Four 12\,m-diameter telescopes were operating from December 2003 and a 5th telescope of 28\,m-diameter became operational in September 2012. In our analysis, the data of 3-4 of the 12\,m telescopes has been utilised. This generally results in field of view of 5$^{\circ}$, an angular resolution (68$\%$ containment radius) of $\sim$ 0.1$^{\circ}$, an energy threshold of $\sim$ 100\,GeV and an energy resolution of $\sim$ 15\% (\cite{Hess:2006}). The SNe in our sample all occur within $\sim$ 2.5$^{\circ}$ of the centre of the field of view.

\subsection{Candidate Selection}\label{sec:Candidates}

\begin{figure}
\begin{center}
\includegraphics[width=1.1\textwidth,trim={1.2cm 1.5cm 0 2.0cm},clip]{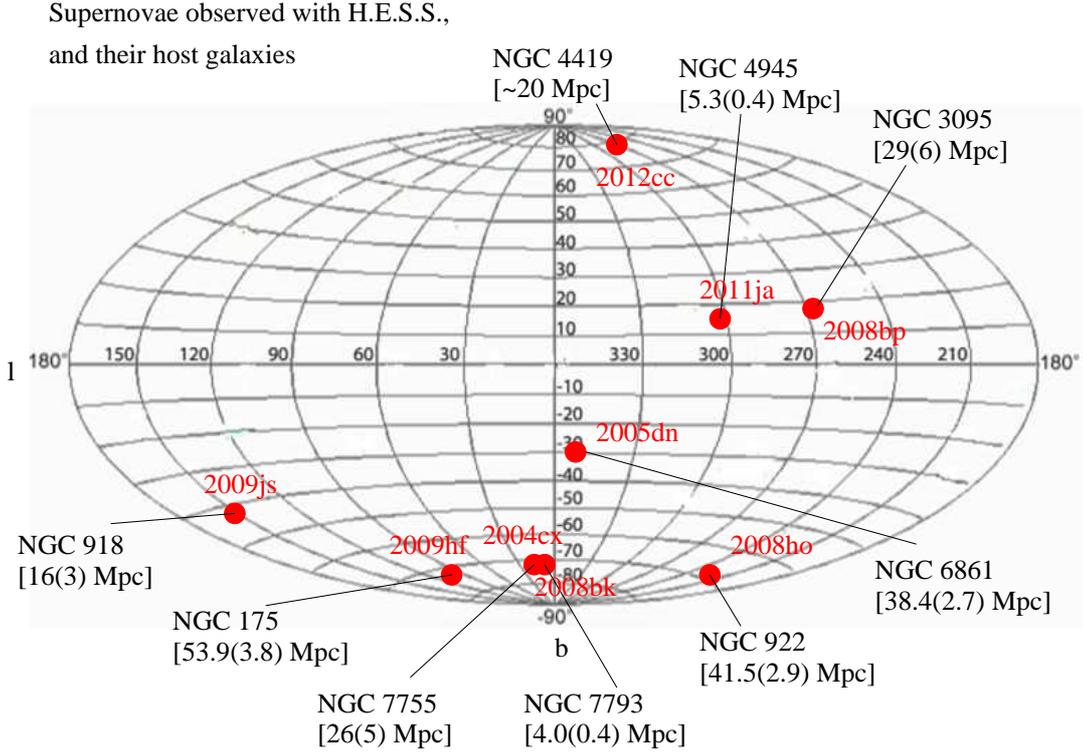}
\caption{Map of the SN candidates selected via the process described in section \ref{sec:Candidates}. Host galaxies and the estimated distances are shown for each SN. The uncertainties in distances are in brackets.}
\label{fig:SNeMap}
\end{center}
\end{figure}

We compared the directions and times of H.E.S.S telescope observations carried out between December 2003 and the 31st of December 2014, with an extensive list of SN candidates compiled from the online IAU\footnote{International Astronomical Union} Central Bureau of Astronomical Telegrams (CBAT) supernova catalogue\footnote{www.cbat.eps.harvard.edu/lists/Supernovae.html}. The SNe sample was constrained to have a host-Galaxy redshift z <0.01 according to the NASA/IPAC\footnote{Infrared Processing $$\&$$ Analysis Center} Extragalactic Database (NED)\footnote{ned.ipac.caltech.edu/forms/z.html}  to ensure that only nearby SNe were considered. If a host galaxy was not stated for a given SN in the CBAT SN catalogue, the SN was discarded from the short-list. The final cross-referenced list of H.E.S.S-observed SNe was constrained to have observations occurring between -7 days and +365 days since the SNe discovery date. The list was further refined by rejecting SNe type 1a and 1c, which are assumed to not occur inside a high-density CSM, hence are not strong cosmic-ray acceleration candidates. The final sample of SNe with serendipitous H.E.S.S. observations are presented on Figure \ref{fig:SNeMap}.


\subsection{Data Analysis}\label{sec:analysis}
Standard quality cuts were applied to remove bad-quality data from each data set, and the Model analysis outlined in \cite{deNaurois:2009} was employed for each candidate SN. For a given SN, multiple OFF regions were defined using the reflected-region method (\cite{Berge:2007}) to estimate the residual background from gamma-ray-like events to subtract from the events detected in the ON-region (0.07 deg radius), selected on the SN position. 
Results were confirmed using the ImPACT fully independent calibration and analysis chain (\cite{ImPACT}) with the reflected background method and standard cuts.

For each 28-minute observation run which passed the criteria outlined in Section\,\ref{sec:Candidates}, the excess was computed (using  N$_{excess}$=N$_{on}$ - $\alpha$N$_{off}$, with $\alpha$ the correction factor for the OFF area with respect to the ON area). The cumulative statistical significance was established using Equation\,17 of \cite{Li:1983}. 

\subsection{Results}
\begin{figure}
\CenterFloatBoxes
\begin{floatrow}
\ffigbox
  {\includegraphics[width=0.5\textwidth]{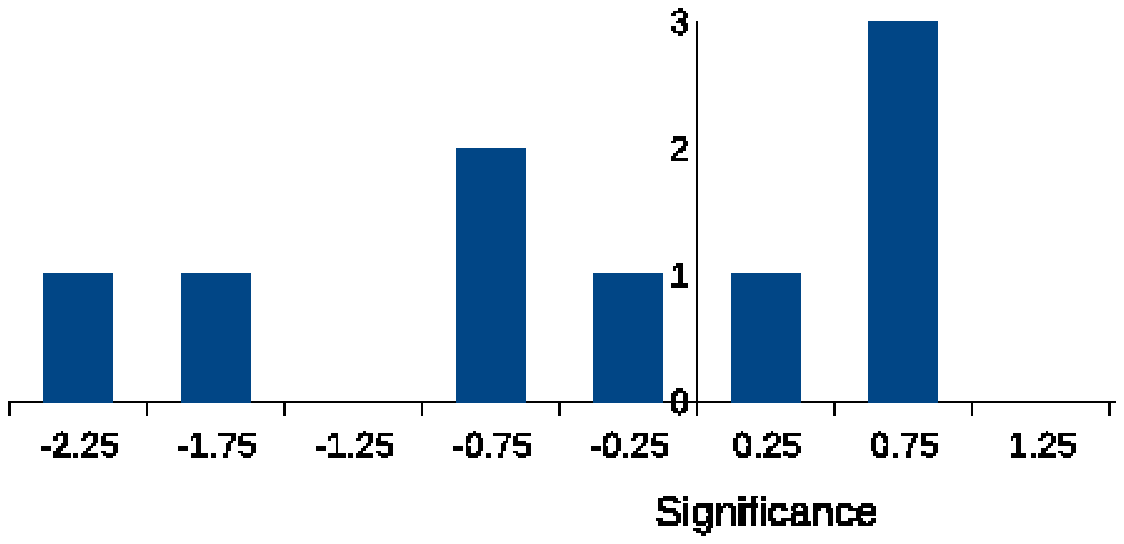}}
  {\caption{Histogram of the $>$1\,TeV gamma-ray significance of the SNe sample.}\label{fig:sig}}
\killfloatstyle
\ttabbox
  {\begin{tabular}{|l|l|c|c|c|c|}
    \hline
   SNe& $E_{Th}$ 			&	UL\scriptsize{$ (>E_{Th})$}			&		UL\scriptsize{$(>1\,\textrm{TeV})$ }\\
&    (TeV)			&	\multicolumn{2}{c|}{(10$^{-13}$\,cm$^{-2}$s$^{-1}$) }\\
    \hline
SN\,2004cx&    	0.18				&	10					&		1.9			\\
SN\,2005dn&    	0.21				&	2.2				&		0.41			\\
 SN\,2008bk&   0.21				&	6.0				&		4.8			\\
SN\,2008bp&    	0.21				&	29					&		5.5			\\
SN\,2008ho&   0.33				&	16					&		7.7			\\
SN\,2009hf&  0.21				&	20					&		5.3			\\
SN\,2009js&   0.63				&	15					&		11			\\
SN\,2011ja&   	0.23				&	13					&		3.9			\\
SN\,2012cc&   	0.72				&	15					&		10			\\
    \hline
  \end{tabular}
  }
  {\caption{Preliminary upper limits (UL) on the integrated flux above the energy threshold, and above 1 TeV (see text).}\label{tab:ul}}
\end{floatrow}
\end{figure}

Figure \ref{fig:SNeMap} displays the observed supernovae along with corresponding host galaxies and distance. Distances range between 4 and 54\,Mpc and discovery dates range between 2004 and 2012. In Table\,\ref{tab:limits} we report the relevant statistics of the gamma-ray observations as described in section \ref{sec:analysis}. Given the serendipitous nature of the observations, the livetime varies between $\sim$1 and $\sim$50\,hours, and the delay between first observation and SN discovery date differs from one SNe to another. The significance distribution, which is represented in Fig\ref{fig:sig}, displays no significant $>$1\,TeV gamma-ray excess for any of the SNe.

Table \ref{tab:ul} shows the TeV gamma-ray upper limits set by H.E.S.S.. These upper limits are derived with 95$\%$ confidence level assuming a power law spectrum of index 2.

\begin{table}
\caption{Observed statistics for each SN event observed with H.E.S.S. (see text).}  \label{tab:limits}
\begin{center}
\begin{tabular}{|l|l|c|c|c|c|c|c|c|}
\hline
SNe		& N$_{on}$ & N$_{off}$ &1/$ \alpha$ & N$_{excess}$		&	Sig 	& { Lifetime} & Delay 			 \\ 		
			& &&&&&(hrs) &  (day) 	\\
\hline
SN\,2004cx	&169 &10387&	65&	  8.7			&		0.7			&39.9&-6	\\
SN\,2005dn	&571 &11452&	19&		-38.6		&		1.5			&53.1&-3		\\
SN\,2008bk	&50 &3652&	54&		-18.0		&		-2.3			&9.6&98	\\
SN\,2008bp	&32 &1860&	60.&		1..1			&		0.2			&4.7&272		\\
SN\,2008ho	&9 &369	&	33&	 	-2.3			&		-0.7			&1.4&36	\\
SN\,2009hf	&43 &1404&35&		3.3			&		0.5			&4.0&0\\
SN\,2009js 	&14 &711&	67&			3.4		&			1		&4.8&94	\\
SN\,2011ja	    &56 &834& 15&		-1.2		&		-0.2			&3.8&91	\\
SN\,2012cc 	&7&660&	74&				-1.9		&		-0.7			&3.0&53	\\
\hline
\end{tabular}
\end{center}
 
{\scriptsize Notes:
 Sig stands for Significance and Delay gives the number of days between first observation and SN discovery.}

\end{table}
 
For all the SNe, the fluxes values are consistent with zero within the sensitivity limits of our observations. No significant TeV gamma-ray emission is found towards any of the SNe within one year of the initial explosion.


\section{Conclusion}

We search for TeV gamma-ray emission from supernovae serendipidously observed with the H.E.S.S. gamma-ray telescope, amending previous efforts (\cite{Lennarz:2013}). 

We find no significant detection above $>$1TeV, and this result completes other recent non detections : a study concerning a large sample of type IIn SNe unseen by Fermi-LAT at GeV energies (\cite{Ackermann:2015}) and the upper limit at TeV energies established by the MAGIC collaboration for the closest Type Ia SN ( \cite{Ahnen:2017}).

The lack of detections reported here do not necessarily indicate that the early phase of SNR development is not conducive to cosmic-ray acceleration, but suggests that only in a subset of the supernovae, those with dense enough CSM, particle acceleration can result in detectable gamma-ray emission.
Furthermore, the incomplete sampling of gamma-ray light curves may have missed features of peak gamma-ray brightness.

Alternatively, gamma-gamma interactions with the photons from the photosphere  may suppress gamma-ray luminosity and the process can be strong at TeV energies, as considered for SN 1993J (see \cite{Marcowith:2014}).

These limits on gamma-ray emission can translate into limits on the physical environment of some of these SN events - something to be investigated in future papers.

\end{document}